\documentclass[12 pt]{article}
\usepackage{graphicx}
\usepackage{amssymb}
\usepackage{amsmath}
\usepackage{amsthm}
\usepackage{mathtools}

\DeclarePairedDelimiter\floor{\lfloor}{\rfloor}

\begin{document}

\title{Non Intrusive Load Monitoring in Chaotic Switching Networks}

{\tiny
\author{P. Garc\'{\i}a\footnote{Permanent address: Laboratorio de Sistemas Complejos, Departamento de F\'{\i}sica Aplicada, Facultad de Ingenier\'{\i}a, Universidad Central de Venezuela,
Caracas, Venezuela.}, X. Dom\'{\i}nguez and D.Chiza\\
Facultad de Ingenier\'{\i}a en Ciencias Aplicadas,\\
 Universidad T\'ecnica del Norte,
Ibarra, Ecuador.}
}

\date{}

\maketitle
\thispagestyle{empty}

\begin{abstract}
In this work, a non intrusive load disaggregation scheme  is proposed. By using a kernel based nonlinear regression strategy, the switching dynamic of an electric network, simulated as a set of RLC circuits with chaotic switching, is approximated using a time series of the total power consumption. The results suggest that the employed methodology can be useful in the design of efficient load disaggregation schemes. 
\end{abstract}

\section{Introduction}
In order to implement energy efficiency strategies in a power system, detailed information on the consumption behavior of the different loads within the system is firstly required.  To perform such load monitoring, there are basically two approaches, intrusive (ILM) and non-intrusive (NILM) methods. The intrusive approach requires individual measurements at every load of interest. Therefore, the need of numerous sets of sensors in ILM systems makes it expensive, complex in installation and more difficult to maintain. Furthermore, the use of smart appliances in ILM systems is common for the load monitoring. 

On the other hand, in the NILM approach, intrusion into the individual or group of appliances when monitoring their power consumption, is not needed. Rather, the load identification is based on the analysis of voltage and current waveforms measured at the electrical service entrance, from which the nature, operating conditions and power consumption, of each load, can be inferred \cite{Lee}. 

When the NILM is applied, the load signatures can be classified into three categories \cite{Abubakar}: steady state signatures, transient signatures and non-traditional signatures. Steady state signatures are the load features extracted when the appliance is on its steady state operation, which include steady state real power, reactive power, rms current, rms voltage, power factor and harmonics, see for instance \cite{Banerjee}. Contrary, the transient state of an appliance refers to the time period between the off state and the steady state operation of the device and vice versa, which is the result of a sudden change in the circuit. This state, can be an on transient or an off transient depending on the changes in the load condition. The transient pattern of most electrical appliances is distinct, which makes them suitable for load identification \cite{Duarte}. Non-traditional signatures such as temperature, light sensing and time of day, can also be used to improve the identification.
As the NILM method employs only one set of sensors at the utility entry, it involves the use of identification algorithms on the collected signatures before the connected loads can be recognized. Within the set of strategies to perform NILM, there are several methods from the computational intelligence field, such as: Back Propagation Artificial Neural Network (ANN) with Artificial Immune Algorithm \cite{Tsai}, Multilayer Feed Forward ANN \cite{Racines}, ANN with genetic programming \cite{Chang} and classification with Neural Network Classifier and Bayes Classifier \cite{Semwal}. 

In any of the above mentioned strategies, the activity recognition in appliance monitoring can have different levels of granularity depending on the application \cite{Ridi} such as the recognition of daily living pattern activities, occupancy detection and user appliance correlation. This information is useful as it permits the inference of underlying patterns which are useful when designing a high-performance load monitoring system.

Regardless of the employed methodology, the common goal of NILM systems consists in providing a fine-grained energy decomposition per appliances in order to procure the optimization of the energy consumption while preserving the comfort of the users. 

In this work, a non-intrusive load disaggregation scheme, by using a nonlinear modeling strategy based on the kernel-Adaline algorithm, is proposed. To achieve this goal, the aim of this work is twofold. First, a simple model of a power network, simulated as RLC branches that are connected and disconnected through chaotic witching, that is, with a disordered but deterministic dynamic, is proposed. Second, an approach to transform the NILM problem into a non-linear regression problem, whose solution is approximate using the kernel-Adaline algorithm \cite{Frieb}, is presented. Although the first objective is straightforward to fulfill, the idea developed in the second, as far as we know, is unpublished. Even though the number of loads used in the simulation is small, the proposed methodology can be scaled to bigger systems.

\section{A simple model for electric loads network.}
In \cite{Hart}, four appliance models have been suggested by the MIT group in order to develop identification algorithms: on/off, finite state machines, continuously variable and permanent state devices. In this work, as a first approach, on/off appliances have been considered. Nevertheless, with a few modifications, the exposed tactic could be extended to other type of loads. 
To simulate an electric network consisting of on/off loads, a simple switched RLC circuit is proposed in Figure 1.

\begin{figure}[h]
\begin{center}
\includegraphics[height=4.0cm]{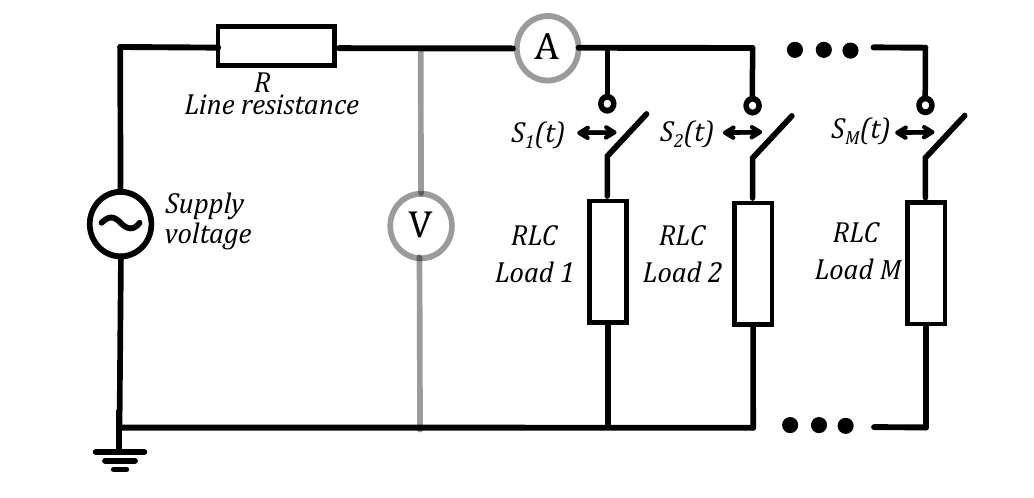}
\caption{Electric network. Every load is represented by a RLC circuit, whose values of R, L and C can be set in order to simulate a particular appliance.}
\label{f1}
\end{center}
\end{figure}

This network in turn can be represented as a system of coupled differential equations:

{ \small
\begin{eqnarray}
R \frac{dq}{dt} +   L_1 \frac{d^2q_1}{dt^2}+ R_1 S_1(t) \frac{dq_1}{dt}  + \frac{q_1}{C_1} & = & \epsilon \cos(wt), \nonumber \\
R \frac{dq}{dt} +   L_2 \frac{d^2 q_2}{dt^2}+ R_2 S_2(t) \frac{dq_2}{dt} + \frac{q_2}{C_2} & = & \epsilon \cos(wt), \nonumber \\
& \vdots & \nonumber \\
R \frac{dq}{dt} +  L_M \frac{d^2q_M}{dt^2}+ R_M S_M(t) \frac{dq_M}{dt} + \frac{q_M}{C_M} & = & \epsilon \cos(wt), \nonumber 
\end{eqnarray}
}

\noindent
where, $q=q_1+q_2+ q_3+ \cdots + q_M$. 

The switching $S_j(t)$, with $j=1, \dots, M$, is performed with a chaotic dynamic, in the sense that the dynamic is deterministic but disordered.
Here, the implementation of $S(t)$ begins by converting the continuous states form the chaotic Logistic map $x_{n+1}= 4 x_n (1-x_n)$, into a chaotic  binary signal as follows:
$$
r_n = \left \{
 \begin{array}{cl}
 K & x_n \leq 1/2 \\
 1 & x_n  > 1/2   
 \end{array} \right.
$$

The continuous switching functions $S_j(t)$ are finally constructed, defining the state of each switch as $S_j(t)=r_n$, with  $n =  \floor{\frac{t}{\tau_j}}$, i. e. the entire part of $\frac{t}{\tau_j}$ and where $\tau_j$  is a set of parameters that dilates the binary signals $S_n$ and provides flexibility to the model, allowing to vary the frequency of the switching but keeping its disordered character. Here, $K=10^5$ is a constant large enough to simulate the switch in OFF or an appliance in standby.

 In this way the time series $x_n$  with discrete time and continuous states is turned into a function of binary states with chaotic evolution and continuous time.  

The resulting system of differential equations is numerically integrated, to later  estimate the rms value of the total current, as a function of time. The results are shown in Figure 2.

\begin{figure}[h]
\begin{center}
\includegraphics[height=5cm]{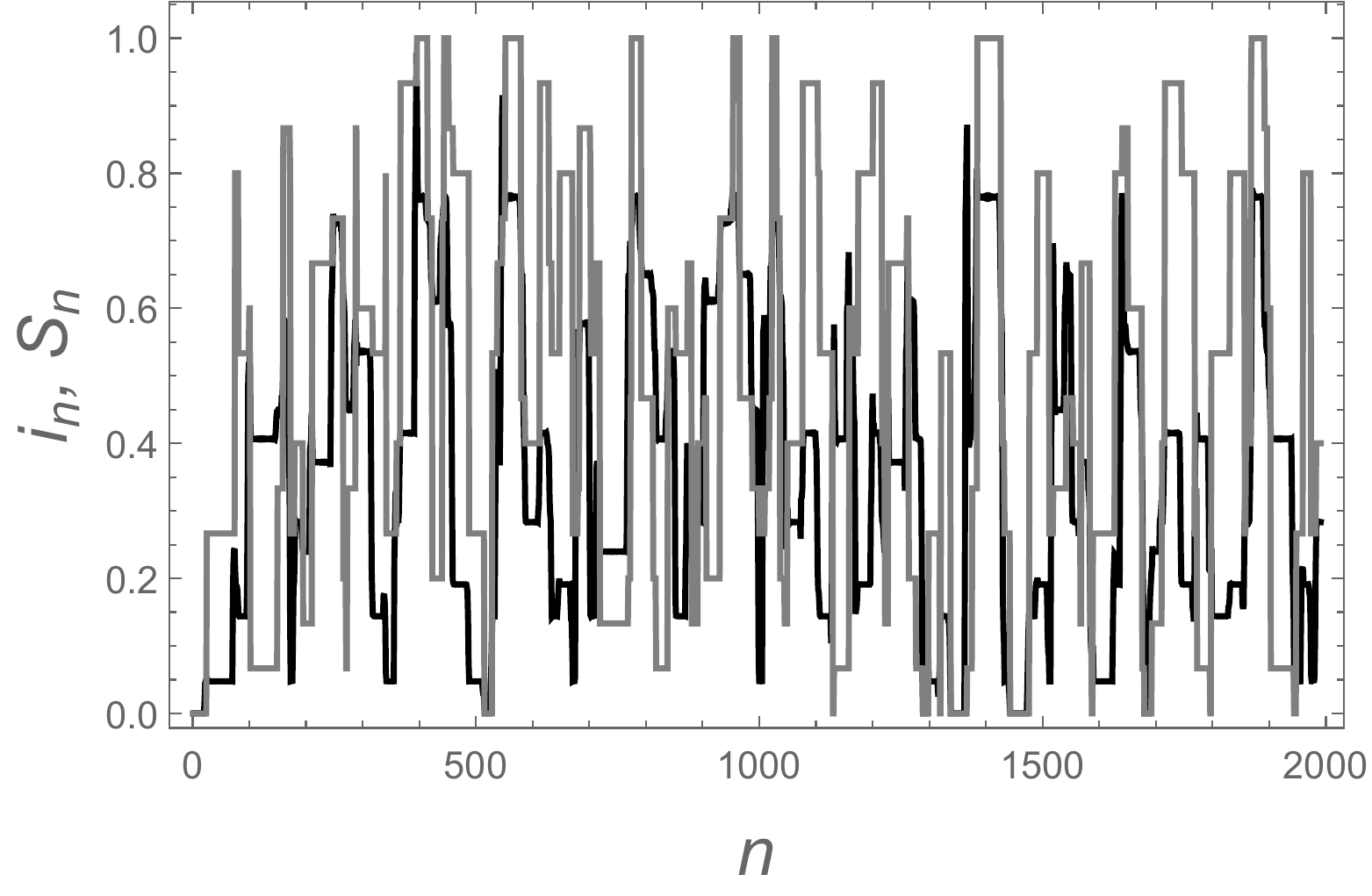}
\caption{Figure shows the total current and switching dynamics. The black and gray  lines represents, the rms value of the total current and the switching states, respectively.}
\label{f3}
\end{center}
\end{figure}

\section{NILM as a nonlinear regression problem}
In this section we attempt to state the problem of non intrusive load monitoring from the point of view of dynamical systems theory. In this approach, the observation of the systems is frequently performed by monitoring the evolution of some of the state variables, namely $x(t)$, taken at uniformly spaced times. The set of all allowed states $x(t)$ is called state space $A$. 

Thus, if our purpose is to predict the future evolution of some dynamical system, $x_{n+1}=f(x_n)$ with $x \in R^m$, then the analysis using embedding methods is a popular alternative \cite{Farmer,Garcia-a}. In these methods, data is organized in {\it delay} vectors ${\bf x}_n = (x_n , x_{n-1} , \cdots, x_ {n-d+1})$. Taken's theorem \cite{Takens} shows that under general conditions, if $d = 2m+1$,  there is a difeomorphism $\Phi : A \to R^d$ such that the dynamical rule governing the evolution of the reconstructed states ${\bf x}_n$ is given by ${\bf x}_{n+1} = \Phi \circ f \circ \Phi^{-1} ({\bf x}_n)$, this is, 

$$
x_{n+1} = F({\bf x}_n). 
$$

An approximation of $F$ can be constructed from the observations $x_n$ and used to forecast the evolution of the system.

In our case, we are dealing with a succession $\{ i_j \}$, $j=1,2,\dots, N$,  of $N$ values of the total current measured at uniform intervals and $\{ S_j^k \}$, with $j=1, 2,\dots, N$ and $k=1,2,\dots, M$, the time series of the states of the $k$-th switch in the time $j$. 

If we re-code the state of the set of switches as a real number:

\begin{equation}
s_j =\frac{ i_{max}}{M} \sum_{k=0}^{M} 2^{k-1} S_j^k,
\label{Switch}
\end{equation}

\noindent
where $i_{max}$ is the maximum value of the total current,  the problem of non intrusive load monitoring can be transformed in the determination of an approach for the dynamical rule $F$, in the form:

$$
\tilde s_j = F({\bf \omega}, {\bf i}_j),
$$ 

\noindent
where the {\it in advance} vectors ${\bf i}_j$ are constructed as ${\bf i}_j = (i_j, i_{j+1}, \cdots, i_{j+d-1})$, $d$ is a parameter related to the influence  of the present state of the switch $s_j$ in  the future of $i_j$   and $\bf \omega$ is a vector of parameters. This function provides the relation between the state of the switches on the present and future values of the total current. Naturally, an approximation of this function allows to estimate the activity of individual loads given the actual value of the total current.

This strategy differs from the usual one in the theory of dynamical systems, in the construction of the state vectors. In this case, instead of constructing vectors with the current state and their predecessors, the state vectors, which we have called {\it in advance} vectors, are constructed with the current state and future values of the variable. This is a consecuence of the causal relationship between $s_n$ and $i_n$.

An efficient alternative to approximate $F$, is the  kernel regression method. In this strategy  the data is mapped to an hight dimensional space  by $\varphi(.)$ and $F$ is fitted by a linear combination of the  $\varphi({\bf i}_j)$  \cite{Garcia-b}. Through the use of reproductive kernels of Hilbert spaces, $K(\cdot, \cdot)$,  it is possible to perform the mapping and its inverse, implicitly. So that:
 
\begin{equation}
F({\bf \omega}, {\bf i}_j)  = \sum_{k=1}^N w_k K({\bf i}_j,{\bf i}_k) + w_{0}.
\end{equation}

In our case, we use the Gaussian kernel

$$K({\bf u}, {\bf v}) = e^{-\frac{\parallel {\bf u} - {\bf v} \parallel^2}{2 p^2}},$$

\noindent
to perform the mapping and its inverse, being $p$ the width of a Gaussian function centered in every sample. 

The solution, to the linear problem that now appears, i. e. the estimation of the parameters $w$, can be approximated using the kernel-Adaline algorithm\cite{Frieb}. This algorithm  is a generalization of the linear Adaline\cite{Widrow}  that allows to approximate non-linear functional relationships, but using strategies for the solution of linear problems. This  algorithm minimizes the least mean squared cost function and additionally, it has the advantage of being numerically robust and  conceptually simple.

\section{Results}
In order to expose the performance of the designed strategy,  a network of $4$ RLC loads with a chaotic switching, is simulated.  In this case, we have generated a total of $2.0 \times 10 ^ 3 $ data. The data, is divided into two segments: one of $1,350$ points, which will be used to train Adaline and another $650$ points intended for the validation of the approach.

The presentation of results begins by showing, in Figure 3, the convergence of the parameters of the kernel-Adaline  $\omega_0$  and one $\omega_i$, randomly chosen ($\omega_r$),  in the case of the data from the training or modeling segment.

\begin{figure}
\begin{center}
\includegraphics[height=5cm]{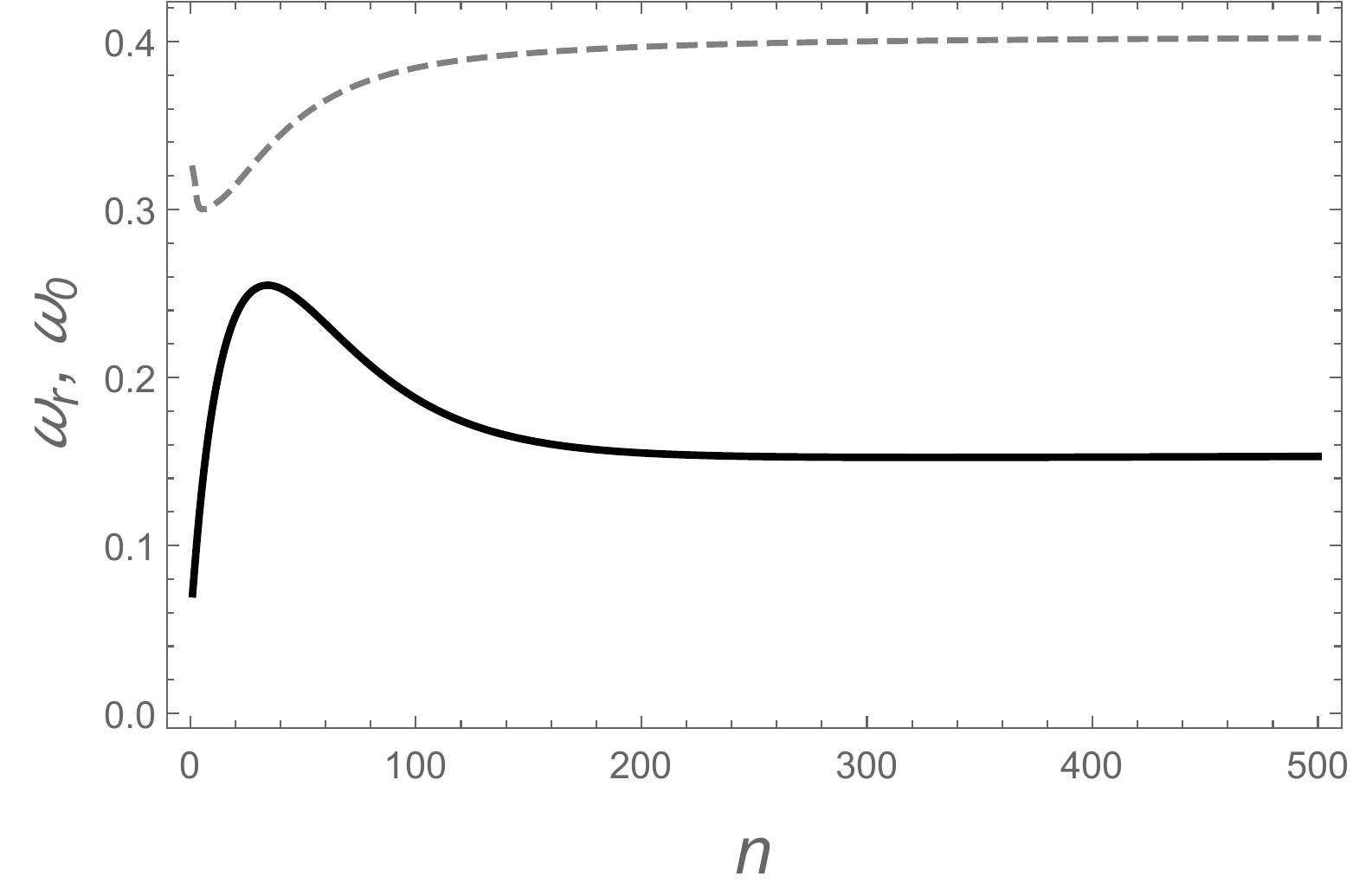}
\caption{Adaline training. The figure shows the evolution of the parameter $\omega_0$ and  $\omega_i$, with $i$ randomly chosen.}
\label{f2}
\end{center}
\end{figure}

Figure 4, shows the performance of the strategy, i.e., it illustrate the quality of the predictions made with the proposed model, in the case of the data corresponding to the training segment. Here, the continuous line shows the evolution of the switching represented as in (1), and the dashed line shows the modeling using (2), when $d = 8$. It is easy to note that, this dimension is dependent on the sampling rate, the step of integration in this case, and from the subsampling rate. The study of these relations, are beyond the scope of this investigation.

Within this framework, the mean squared error of the switching state  prediction, defined as:

$$
E = \frac{1}{N} \sum_{i=1}^N (s_n - \tilde s_n)^2,
$$

\noindent
from the total current, is  $E=0.0023$. Figure 5 graphically shows this result, in the case of the data corresponding to the validation segment.

\begin{figure}
\begin{center}
\includegraphics[height=5cm]{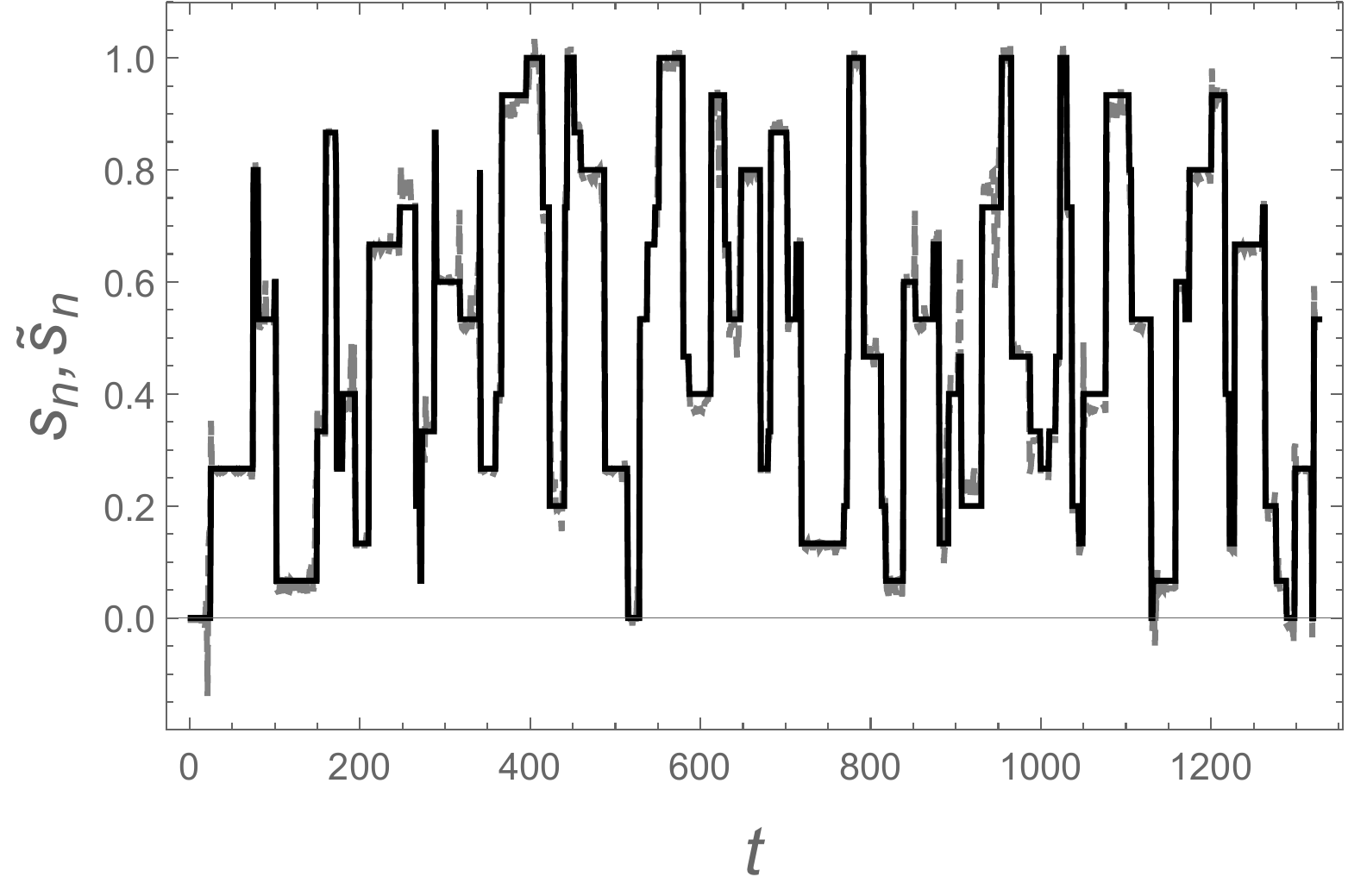}
\caption{Figure shows the switching dynamics. The continuous and dashed  line represents, the switching dynamic and it approximation, respectively, in the case of the training segment.}
\label{f4}
\end{center}
\end{figure}

\begin{figure}
\begin{center}
\includegraphics[height=5cm]{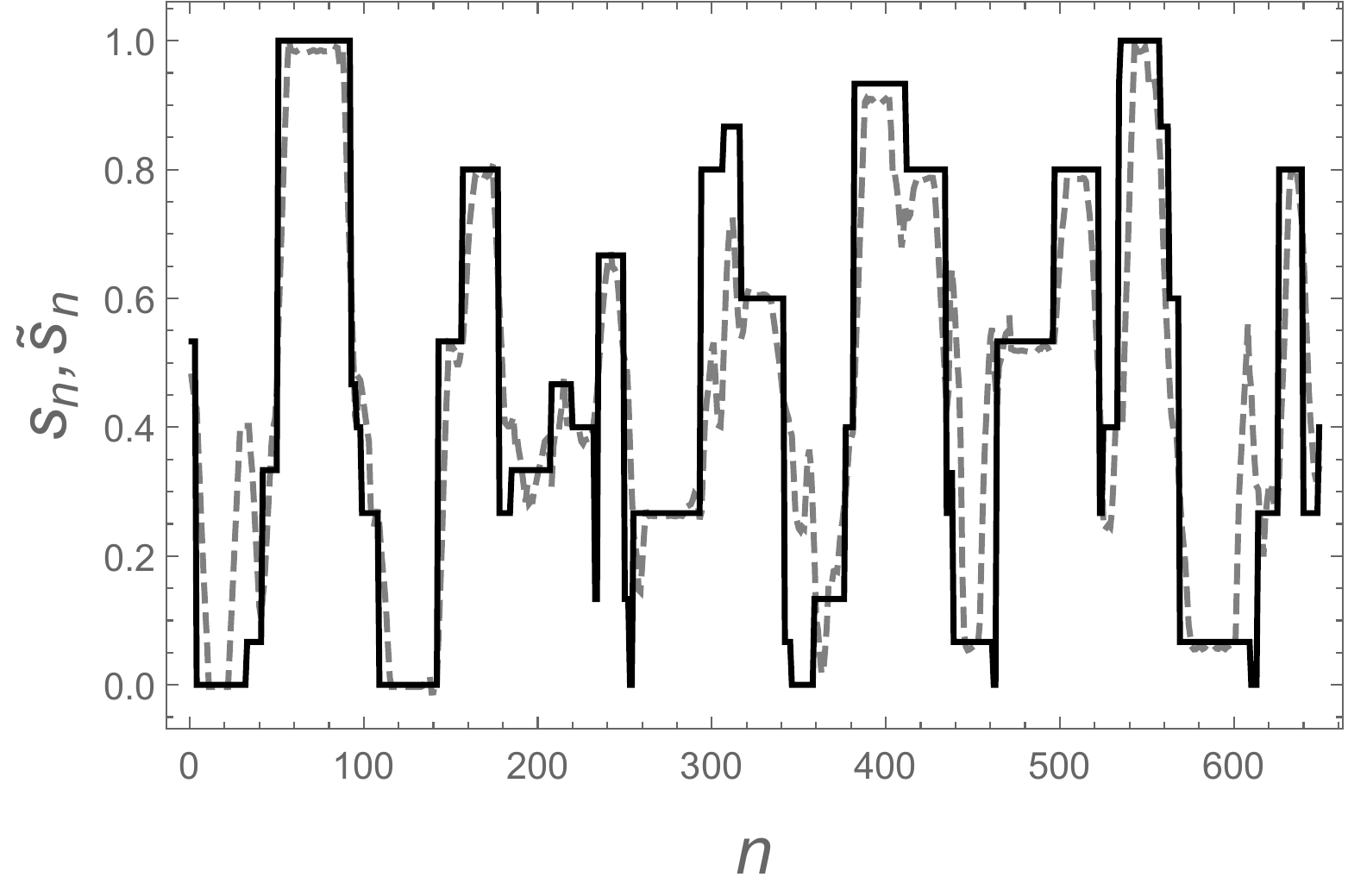}
\caption{Figure shows the switching dynamics. The continuous and dashed  lines represents, the switching dynamic and it approximation, respectively, in the case of the validation segment.}
\label{f4}
\end{center}
\end{figure}

\section{Concluding remarks}

A NILM scheme, based on nonlinear regression  strategy and implemented by mean of the  kernel-Adaline algorithm,  has been presented. 
This representation of NILM, as a non-linear regression problem using kernel regression strategies, as far as we know, represents a novelty in the treatment of the problem. 

The numerical experiments suggests that although the forecast quality is appropiate, the quality of the disaggregation, using this method, depends on the parameters $\tau_i$. This represents an interesting theoretical and computational problem which guides future research. 

Due the characteriristics of the proposed representation of the problem, the extension to the case of networks with loads of another type, or with more elements, could be addressed in a simple manner.

Finally, it is worth to mentioning that the NILM problem is highly interesting, not only due to its practical applications in the efficient energy consumption, but also from the point of view of dynamical systems theory. In this field, it represents a complex problem of coupled linear systems of differential  equations, but not autonomous because to a complex switching functions.

\end{document}